\documentclass[11pt,a4paper,portrait]{article}

\usepackage{natbib}
\usepackage{graphicx}

\begin{document}

\title{\textbf{Which solar EUV indices are best for reconstructing the solar EUV irradiance ? }}           
\author{T. Dudok de Wit$^1$ \and M. Kretzschmar$^2$ \and J. Aboudarham$^3$ \and P.-O. Amblard$^4$ \and F. Auch\`ere $^5$ \and J. Lilensten $^6$}

\date{{\small\today -- to appear in Adv. Space. Research 2007}}

\maketitle                  

\begin{center}
{\small
$^1$ LPCE, CNRS and University of Orl\'eans, 3A av. de la Rech. Scientifique, 45071 Orl\'eans, France \\ 
$^2$ SIDC, Royal Observatory of Belgium, Ringlaan 3, 1180 Brussel, Belgium \\
$^3$ LESIA, Paris Observatory, 5 Place Jules Janssen, 92195 Meudon, France \\
$^4$ LIS, CNRS, 961 Rue de la Houille Blanche, BP 46, 38402 St. Martin d'H\`eres, France \\
$^5$ IAS, CNRS and University of Paris-Sud, B\^atiment 121, 91405 Orsay, France \\
$^6$ LPG, CNRS and Joseph Fourier University, B\^atiment D de Physique, BP 53, 38041 Saint-Martin d'H\`eres, France
}
\end{center}

\begin{abstract}
The solar EUV irradiance is of key importance for space weather. Most of the time, however, surrogate quantities such as EUV indices have to be used by lack of continuous and spectrally resolved measurements of the irradiance. The ability of such proxies to reproduce the irradiance from different solar atmospheric layers is usually investigated by comparing patterns of temporal correlations. We consider instead a statistical approach. The TIMED/SEE experiment, which has been continuously operating since Feb. 2002, allows for the first time to compare in a statistical manner the EUV spectral irradiance to five  EUV proxies: the sunspot number, the f10.7, Ca K, and Mg II indices, and the He I equivalent width.

Using multivariate statistical methods, we represent in a single graph the measure of relatedness between these indices and various strong spectral lines. The ability of each index to reproduce the EUV irradiance is discussed; it is shown why so few lines can be effectively reconstructed from them. All indices exhibit comparable performance, apart from the sunspot number, which is the least appropriate. No single index can satisfactorily describe both the level of variability on time scales beyond 27 days, and relative changes of irradiance on shorter time scales.   
\end{abstract}


\section{Introduction}

The solar irradiance in the EUV range\footnote{It is common practice in aeronomy to distinguish XUV (1-30 nm), EUV (30-121 nm) and FUV (122-420 nm). We shall use the generic term EUV for all of them.} is a key parameter for aeronomy \citep{hinteregger81} and for space weather \citep{lathuillere02}. It is also one of the least accessible parameters, as EUV measurements must be carried out above the terrestrial atmosphere. Moreover, space-borne EUV detectors suffer from instrument degradation. Not surprisingly, there have been very few continuous and spectrally-resolved measurements in the spectral range that is of interest for aeronomy, typically between 20 and 150 nm. This situation has led to the widespread use of proxies as substitutes for the irradiance \citep{cebula98, tobiska00, lathuillere02}.

Many studies have been devoted to the comparison between solar EUV proxies and the solar EUV irradiance; \citet{floyd05} have recently reviewed three decades of results. The physical connection between these indices and the irradiance, however, is at best indirect, and there are also substantial differences in the way these different quantities are measured. In spite of this, most proxies reproduce the variability of the EUV irradiance remarkably well. Their strong temporal correlations emerge as a result of close connections between the irradiance mechanisms at different solar atmospheric layers, and yet significant discrepancies persist. The accurate reconstruction of the solar EUV irradiance from surrogate quantities remains an open problem.

Most solar irradiance studies are based on detailed comparisons of events. Differences in the time evolution indeed provide direct insight into the underlying physics. We consider here a different and novel approach that uses a global representation and shows how the EUV irradiance and the proxies are related to each other. This statistical approach was recently introduced with the aim to determine how the solar spectral irradiance could be reconstructed from the linear combination of a few (typically 4 to 8) spectral lines \citep{ddw05}, following an earlier investigation based on physical criteria \citep{kretzschmar04}. Here we use the same approach to compare several EUV indices with a selection of strong EUV lines. Using two different normalisations, we investigate how well each index reproduces emissions that originate from different layers of the solar atmosphere, and which combination of indices would be appropriate.


\section{Solar indices for the EUV spectral irradiance}

The lack of continuous observations in the EUV is a long-standing problem in solar irradiance studies. This situation first improved in 1991 with the continuous irradiance measurements from the SOLSTICE and SUSIM instruments \citep{rottman00}. A second major improvement came from the EUV spectrometer onboard the TIMED satellite \citep{woods05}. With several years of continuous operation, this instrument for the first time allows the EUV spectral irradiance to be compared statistically against EUV indices. The EVE instruments onboard the future Solar Dynamics Observatory will soon provide additional spectral resolution and coverage.

In this study, we make use of four years of daily TIMED data. Although four years is not sufficient for properly validating the impact of the solar cycle, it already provides interesting insight. Our time interval (Feb. 2002 until May 2006) starts shortly after solar maximum, and includes the full decay of the cycle down to solar minimum.

The six quantities we consider here are:

\begin{description}

\item[The spectral irradiance] measured by the Solar Extreme Ultraviolet Experiment (SEE) \citep{woods05} onboard TIMED. We consider daily-averaged solar spectral irradiance measurements made by EUV grating spectrograph that is part of SEE. This spectrograph covers 26 to 194 nm with 0.4 nm spectral resolution; its measurements are corrected for atmospheric absorption and instrument degradation, and are normalised to 1 AU. TIMED makes several measurements per day and so the contribution of solar flares is subtracted. Our analysis is based on version 8 data.

We focus here on daily intensities of 38 strong spectral lines, from February 8th, 2002 until May 14th, 2006. Although more recent SEE data exist, our time span is constrained by the availability of the indices. The 38 spectral lines are shown in Fig.~\ref{fig_spectrum}.

\begin{figure}[!htb]
\begin{center}
\includegraphics[width=13.5cm]{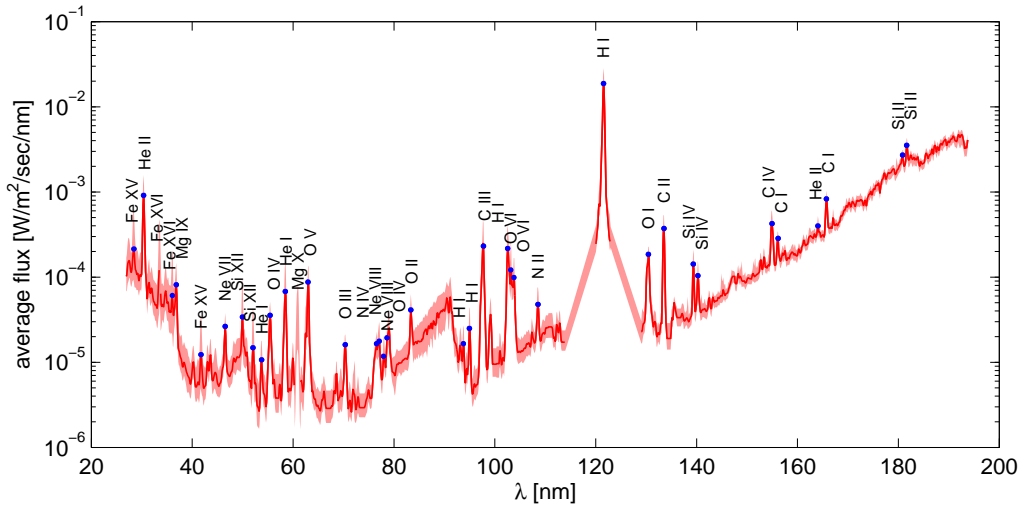}
\caption{Time-averaged EUV spectrum from SEE. The shaded area expresses the variability of the irradiance between Feb. 2002 and May 2006. The 38 spectral lines of our set are shown. A filter blocks out the wings of the bright H I Lyman-$\alpha$ emission.}
\label{fig_spectrum}
\end{center}
\end{figure}

\medskip

\item[The international sunspot number,] as computed daily by the Royal Observatory of Belgium. This oldest and best known gauge of solar activity is connected to the EUV irradiance through the presence around sunspots of hot plages and faculae, in which the EUV emission is enhanced. Short-term variations of the sunspot number, however, do not always correlate well with the EUV irradiance \citep{donnelly86}. This quantity cannot properly capture features such as centre-to-limb variations and the emission of decaying sunspots.

\medskip

\item[The decimetric f10.7 index] which is a daily measurement of the radio flux at 10.7 cm made by the Penticton observatory. This radiometric index measures both thermal emission and electron gyro-resonance emission, which issue from the high chromosphere and low corona. The f10.7 index is today widely preferred to the sunspot number since it is easier to measure from ground and it is better correlated with the EUV irradiance \citep{donnelly83,floyd05}.

\medskip

\item[The Mg II index,] which is the core-to-wing ratio of the Mg II line at 280 nm and probes the high chromosphere. This index, which was first developed by \citet{heath86} has been shown to be an excellent surrogate for the UV irradiance; it also fits the EUV irradiance also quite well \citep{thuillier01,viereck01} in spite of differences in the way the index is derived from various experiments. We use the composite Mg II data set compiled by the Space Environment Center (NOAA).

\medskip

\item[The Ca K index] is the normalised intensity of the Ca II K-line at 393 nm and has been recognised early on as a interesting index for the UV domain \citep{lean82}. Ca K line images are routinely used to track the evolution of plages and the network, whereas the Mg II line is better suited for faculae. The Mg II index is generally considered to be better correlated with chromospheric emissions than the Ca K index \citep{hedin84}. Our Ca K data were compiled by the NSO at Kitt Peak.

\medskip

\item[The equivalent width of the He I 1083 nm infrared absorption line] has been computed from ground-based images since the 1980's \citep{harvey94}. This quantity has been shown to probe the cold contribution of the EUV spectrum quite well \citep{donnelly86}. We use the He I index compiled by the NSO at Kitt Peak.
\end{description}

All quantities, apart from the Mg II index and the EUV irradiance, can be measured from ground. The He I and Ca K indices, however, are susceptible to weather conditions. Several quantities suffer from data gaps and the He I data are only available until Sept. 21, 2003. Data gaps of less than a month can easily be filled by a multivariate variant of the interpolation scheme developed by \citet{kondrashov06}, which performs here remarkably well, owing to the redundancy of the data. We made no attempt, however, to extrapolate the He I data. This restriction does not affect our analysis but it means that the results obtained with the He I index may be biased by lack of sufficient temporal coverage.

All EUV irradiances and indices are strongly correlated for yearly variations, but show significant differences in their short-term variations. Direct visualisation of their time series has so far been the standard way of looking at these differences. To the best of our knowledge, Pearson's correlation coefficient is the only quantity that has been used to quantify similarity. Such statistical measures, however, only reveals how quantities are related pairwise. We shall now show how all the information can be gathered in a single representation.


\section{The method: multidimensional scaling}
\label{sec:method}

The analysis method we advocate here is identical to the one we used for reconstructing the EUV spectrum from a reduced set of spectral lines \citep{ddw05}. We first quantify the connectivity between two observables by means of their Euclidean distance
\begin{equation}
\delta_{kl} = \sqrt{\int \Big(y_k(t) - y_l(t) \Big)^2 dt} , 
\end{equation}
where $y_k(t)$ and $y_l(t)$ are the time series of any of the quantities listed in the preceding section, after some suitable normalisation (to be discussed later). The smaller this distance is, the more related the dynamics of the two quantities is and the more likely their common physical origin is. 

We next represent each quantity by a single point in a multidimensional connectivity map, in which the distance between any pair of points equals the dissimilarity $\delta$. The number of observables is 43 (38 spectral lines + 5 indices), which means that this map should formally have 42 dimensions. Since, however, most quantities are strongly correlated, the dimensionality can be strongly reduced without losing pertinent information. This reduction considerably eases the visualisation and the interpretation. The technique for building such a connectivity map is called \textit{multidimensional scaling} and is well known in the multivariate statistics literature \citep{chatfield90}. Incidentally, since we are using an Euclidean distance, the low-dimensional representation of the connectivity map is nothing but a projection onto the first principal axes of the data. Using principal component analysis, we then express the spectral variability as a linear combination of separable modes \citep{chatfield90}
\begin{equation}
y_k(t) = \sum_{i=1}^N A_i f_i(t) g_i(k) , \qquad k=1,2, \ldots, 43
\end{equation}
with the orthonormality constraint
\begin{equation}
\langle f_i(t) f_j(t) \rangle = \langle g_i(k) g_j(k) \rangle = 
\left\{
\begin{array}{lll}
0 & \textrm{if} & i \neq j \\
1 & \textrm{if} & i = j \\
\end{array}
\right. ,
\end{equation}
where $\langle . \rangle$ means ensemble averaging. The weights are traditionally sorted in decreasing order $A_1 \ge A_2 \ \cdots \ge A_N \ge 0$. The number of modes $N$ here equals the number of observables. Large weights correspond to modes that describe features shared by many observables. Since most quantities exhibit very similar time evolutions, we can expect very few modes to capture the salient features of the full data set. The proportion of variance accounted for by the $i$-th mode is
\begin{equation}
V_i =  \frac{A_i^2}{\sum_{j=1}^N A_j^2} .
\end{equation}
As will be shown below in Sec.~\ref{sec:comparison}, one or two modes only are needed to describe over $90\%$ of the variance. This remarkable result is a direct consequence of the strong connections between solar emission processes at different altitudes.

As shown by \citet{chatfield90}, the coordinates of our observables along the $i$'th axis of the connectivity map are simply given by $A_i g_i(k)$. A two-dimensional map is needed if two modes describe the data. Three or more dimensions are necessary if there are more outstanding modes with large weights. The time-profile $f_i(t)$ associated with the $i$'th axis expresses the type of dynamics that is shared by observables lying along that axis. An inspection of these time-profiles is needed to interpret the axes.


\section{Choice of the normalisation}

The main quantity of interest here is the relative position of our observables on the low-dimensional connectivity map. The multidimensional scaling technique is not scaling invariant and so it is important to specify how the data are normalised. There are essentially two choices:
\begin{enumerate}
\item The default choice in statistics is \textit{standardisation}
\begin{equation}
y_k(t) \longrightarrow \frac{y_k - \bar{y}_k}{\sigma_k}
\end{equation}
in which each quantity is centred with respect to its time average $\bar{y}_k$ and then reduced by its standard deviation $\sigma_k$. By doing so, we put all quantities on equal footing irrespective of their level of temporal variability. Such a normalisation is appropriate for comparing UV indices with hot coronal lines, since the two exhibit very different levels of modulation with solar rotation. 

If two standardised quantities $y_k$ and $y_l$ overlap on the connectivity map, then they are connected by a linear relationship $y_k \approx \alpha y_l + \beta$, where $\alpha$ and $\beta$ are two constants.

\item Another choice (hereafter called \textit{normalisation}) consists in normalising each quantity with respect to its time-average only
\begin{equation}
y_k(t) \longrightarrow \frac{y_k - \bar{y}_k}{\bar{y}_k}
\end{equation}
This choice preserves the information about the level of variability. 

Two normalised quantities $y_k$ and $y_l$  coincide on the connectivity map if they are linearly proportional to each other: $y_k \approx \alpha y_l$, where $\alpha$ is a constant.
\end{enumerate}

\begin{figure}[!htb]
\begin{center}
\includegraphics[width=12cm]{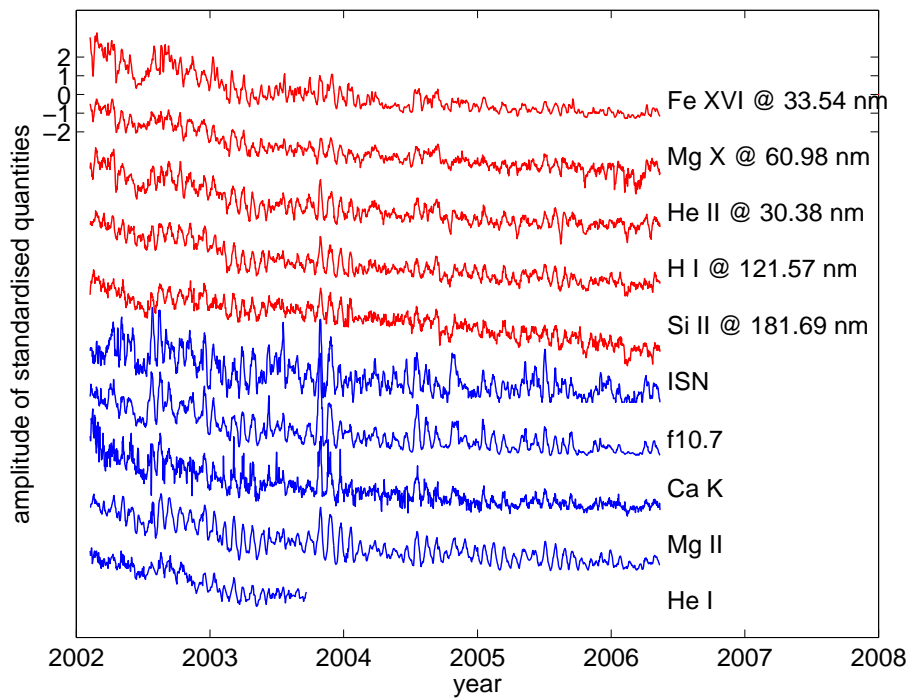}
\caption{Time evolution of five spectral lines and five indices. All quantities are standardised.}
\label{fig_excerptn}
\end{center}
\end{figure}

\begin{figure}[!htb]
\begin{center}
\includegraphics[width=12cm]{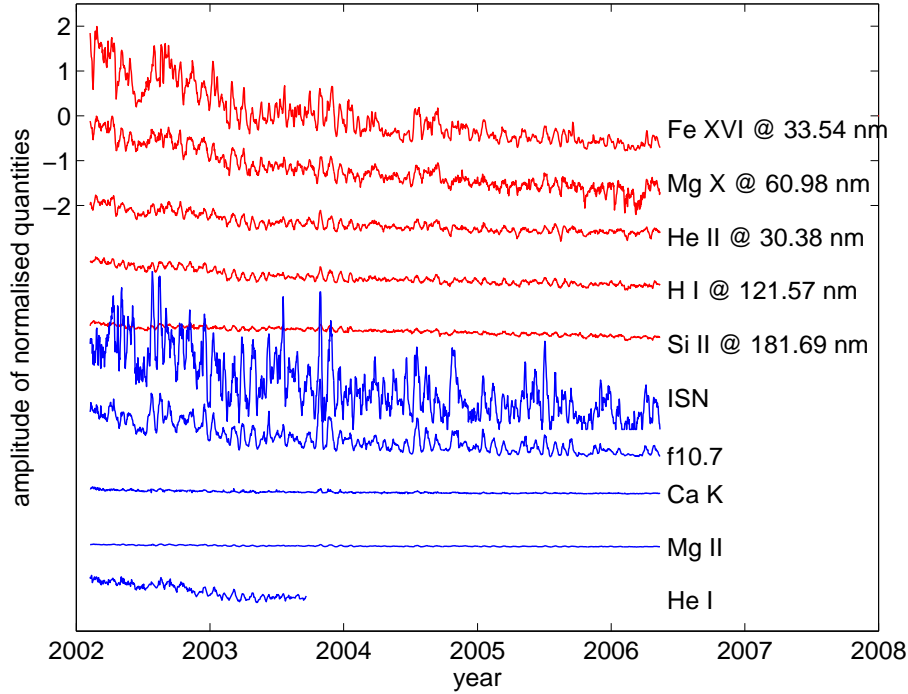}
\caption{Same plot as Fig.~\ref{fig_excerptn}, but with normalised quantities.}
\label{fig_excerpt}
\end{center}
\end{figure}

In principal component analysis, the decomposition of standardised data involves the diagonalisation of the data correlation matrix, whereas for normalised data, the data covariance matrix is diagonalised. The two normalisations are illustrated in Figs.~\ref{fig_excerptn} and \ref{fig_excerpt}, in which we respectively plot standardised and normalised quantities. Five typical spectral lines and the five indices are shown. The most conspicuous features are the decaying solar cycle and the solar rotation, which is responsible for a 27-day modulation. A 13.5-day modulation, which is caused by centre-to-limb variations \citep{crane04}, is sometimes apparent. The stronger variability of hot coronal lines, such as Fe XVI, only comes out in normalised quantities, whereas standardisation is better suited for comparing fine details in the time evolution. 

By representing each observable by a single point on the connectivity map, we gain a global view that provides interesting insight into the connection between the different indices and the EUV irradiances. It must be stressed, however, that differences in long-term trends may be affected by instrumental effects and in particular by sensor degradation \citep[e.g.][]{deland98}. Short-term differences are more likely to be associated with differences in the emission processes \citep{donnelly86, crane01, floyd05}. Unfortunately, all scales are mixed here. A more general method for overcoming this is in preparation \citep{ddw07}.

Caution should also be exercised in extrapolating our results, since our statistical sample covers less than one solar cycle. It has been shown before \citep[e.g.][]{floyd05} that some indices may not agree equally well during different phases of the solar cycle. We checked this by repeating our analysis for shorter sequences. No striking differences were found. Nevertheless, a statistical description like ours formally is complete only once the sample covers at least one full magnetic cycle (22 years).


\section{Comparison between indices and irradiances}
\label{sec:comparison}

Figures \ref{fig_map_norm} and \ref{fig_map_stand} show the connectivity maps obtained respectively with normalised and with standardised data. These two plots contain the main results of our study. We built the connectivity maps using 38 spectral lines from TIMED/SEE and 5 indices. Although their true dimension is 42, two dimensions are sufficient to display the cloud of points while preserving the distances as defined above. The fraction of signal variance that is described by the first four axes or dimensions is listed in Table~\ref{table_fraction}.

\begin{table*}[!htb]
\vspace*{5mm}
\caption{Fraction of the variance of the data that is described by the first four axes of the connectivity map. The axes are conventionally sorted by decreasing variance.}
\label{table_fraction} 
\begin{center}
\begin{tabular}{crr}
\hline
axis & normalised & standardised \\
$i$  & data [\%] $V_i$ & data [\%] $V_i$ \\  \hline
1 & 93.83 & 88.46 \\
2 & 1.56 & 5.31 \\
3 & 1.01 & 0.86 \\
4 & 0.54 & 0.65 \\
\hline 
\end{tabular}
\end{center}
\end{table*}

\subsection{Comparison with normalised data} 

\begin{figure*}[!htb]
\begin{center}
\includegraphics[width=\textwidth]{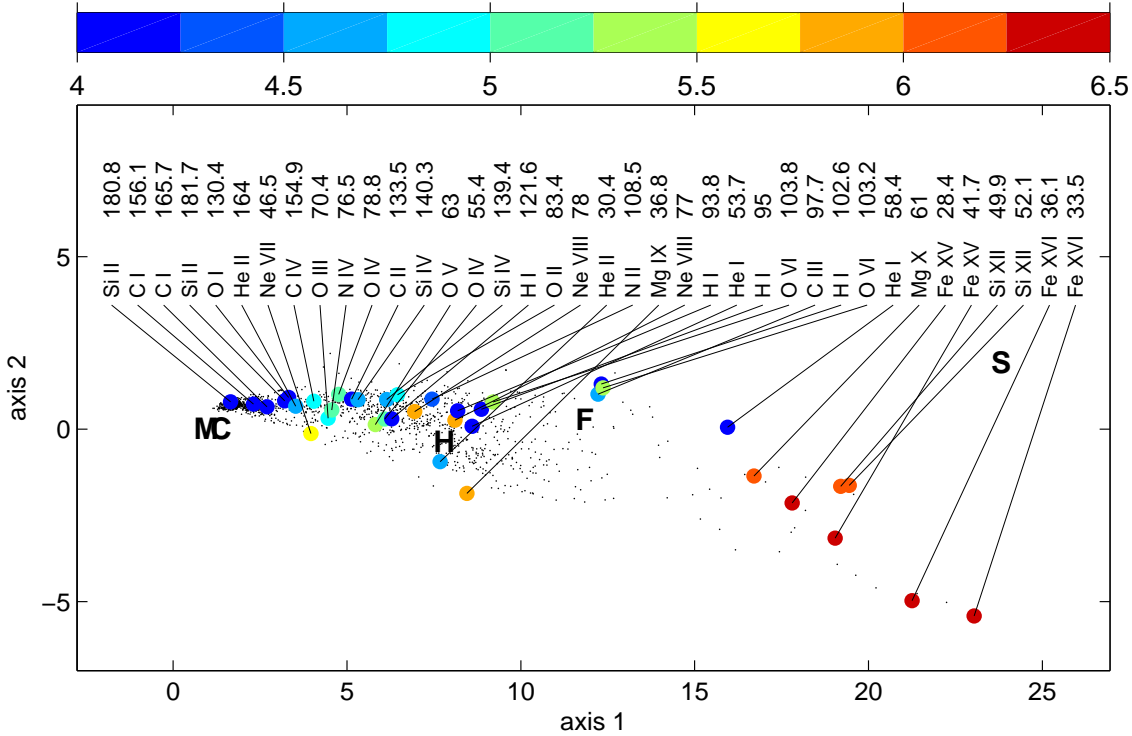}
\caption{Two-dimensional connectivity map, showing 38 normalised spectral lines and 5 normalised indices. The distance between each pair of quantities reflects the degree of dissimilarity in their time evolution. The five indices are labelled by letters: f10.7 index (F), sunspot number (S), He I equivalent width (H), Ca K index (C) and Mg II index (M). Each wavelength of TIMED/SEE is represented by a small dot, except for the 38 strongest spectral lines, whose colour indicates the decimal logarithm of the characteristic emission temperature. The coordinates along each axis are defined in Sec.~\ref{sec:method}.}
\label{fig_map_norm}
\end{center}
\end{figure*}

\begin{figure*}[!htb]
\begin{center}
\includegraphics[width=\textwidth]{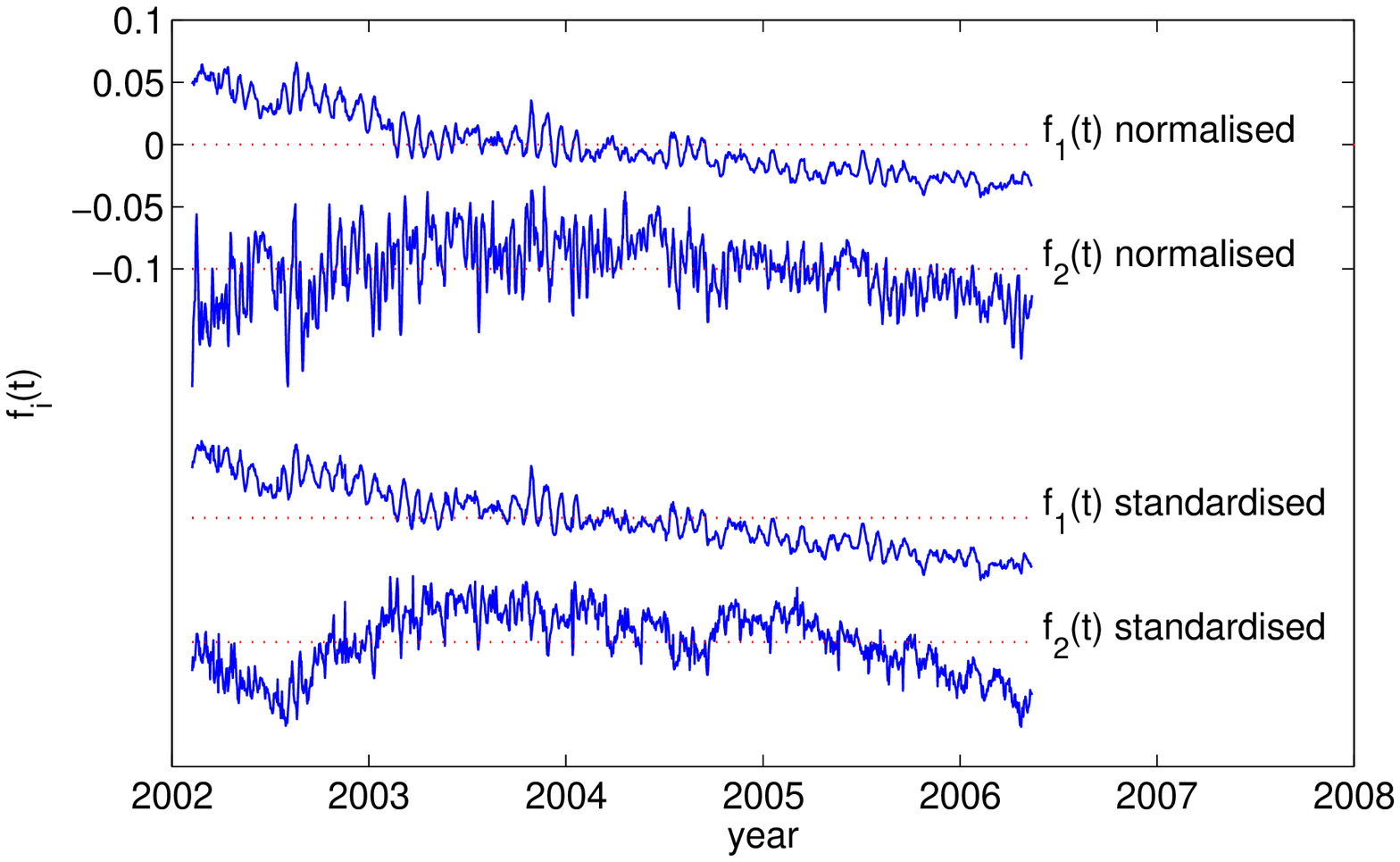}
\caption{Time-profiles associated with the first two axes of the normalised data (top) and the standardised data (bottom). These profiles reveal what kind of dynamics is described by the different axes. The time series have been shifted vertically.}
\label{fig_timeseries}
\end{center}
\end{figure*}

The first axis of the connectivity map (see Fig.~\ref{fig_map_norm}) is by far the most important one, as it describes over 93 \% of the variance. The corresponding time-profile is shown in Fig.~\ref{fig_timeseries}. Not surprisingly, the time-profile captures the decaying solar cycle and the 27-day modulation, which are common to all indices and to all spectral lines.  The second axis captures a slow trend and the 13.5-day modulation that is typical for hot coronal lines. These results have already been discussed by \citet{ddw05}. For standardised data, the second axis only captures slow trends. The third and subsequent axes still contain some variance and so cannot be fully neglected. Their contribution, however, is weak and in contrast to the first two axes, depends on the time interval of the observations. Adding a third dimension would also considerably complicate the visualisation.

The connectivity map obtained with normalised data is very similar to the one published by \citet{ddw05}, using half as many data\footnote{The main difference is in the location of the f10.7 index, which tends to move downward as the solar cycle progresses.}. The rightmost lines or indices are those which are most strongly modulated by the solar cycle and by solar rotation. Hot coronal lines are well known to be strongly modulated \citep{woods05}. Some colder but optically thick lines such as He II (30.4 nm) and He I (53.7 nm) are also modulated, presumably because they are driven by coronal emissions. One of the most conspicuous features is a strong correlation between the horizontal position of the lines (i.e. their degree of modulation) and their emission temperature, with the coldest ones on the left. The only exceptions are the optically thick H and He lines, whose dynamics is clearly different. 

The vertical axis in Fig.~\ref{fig_map_norm} mainly describes the level of 13.5-day modulation, see Fig.~\ref{fig_timeseries}. This axis may therefore be associated with centre-to-limb variations, which peaks as expected for hot coronal lines. Lines that are located at or below the transition region are clustered in the upper left corner, close to the thermal continuum.

The main point of interest in Fig.~\ref{fig_map_norm} is the location of the indices. All of them are located in the vicinity of the cloud of spectral lines, which confirms their validity as surrogates for the EUV irradiance. A closer look, however, reveals significant differences. No single line lies close to the sunspot number, which means that the EUV irradiance cannot be proportional to it. This isolated location of the sunspot number can be explained by its strong sensitivity to solar activity (making it appear on the far right) and its non-radiometric nature (making it insensitive to the geometric location of active regions). The f10.7 index is better located between the coronal lines and the colder ones, and is therefore better suited for EUV studies. 

The Mg II and Ca K indices in comparison appear much closer to the origin, where the chromospheric lines are located. The reason for this is their weak modulation by the solar cycle. These two indices are therefore appropriate for describing the less energetic part of the EUV spectrum. Notice however that both are located slightly outside of the cluster of EUV irradiances, which means that they do not fully describe the latter. 

The He I index has been advocated for transition region lines \citep{donnelly86} and this is indeed clearly confirmed by our plot. The third axis of the connectivity map (not shown), however, reveals a small but significant separation that excludes an exact reproduction of any of the EUV lines. Let us also recall that this index should be interpreted with care by lack of recent data.

Some words of caution are needed regarding the interpretation of the lines. Because of the 0.4 nm spectral resolution of SEE, what we call a spectral line actually is a blend of emission originating from the centre of the line, its wings and often from unresolved nearby lines. This typically happens with the strong He II emission at 30.38 nm, which is polluted in active regions by the weaker Si XI line at 30.32 nm. Because of this, some lines may actually be located slightly off their true position. This effect is hard to quantify without having access to higher resolution data. We nevertheless tested it by artificially mixing some lines. A maximum horizontal displacement of about one unit in Fig.~\ref{fig_map_norm} is not excluded, especially for the hot coronal lines.

A second problem is the degradation of the EUV detectors and the subsequent decrease in the signal-to-noise ratio. This problem is accentuated by the absence of recalibration since the last calibration rocket flight in October 2004. We do indeed see an increase of the scatter of the points in the connectivity maps (especially with standardised data) when the analysis is restricted to more recent intervals. This increased scatter probably also has to do with the lack of variability as compared to earlier years. An indication may be given by the number of modes that are needed to properly reproduce a given spectral line. More modes are needed to describe features that exhibit a more complex dynamics \citep{aubry91}. In our case, this additional complexity may be interpreted as a departure from redundancy, i.e. an increased noise level.

\subsection{Comparison with standardised data} 

\begin{figure*}[!htb]
\begin{center}
\includegraphics[width=\textwidth]{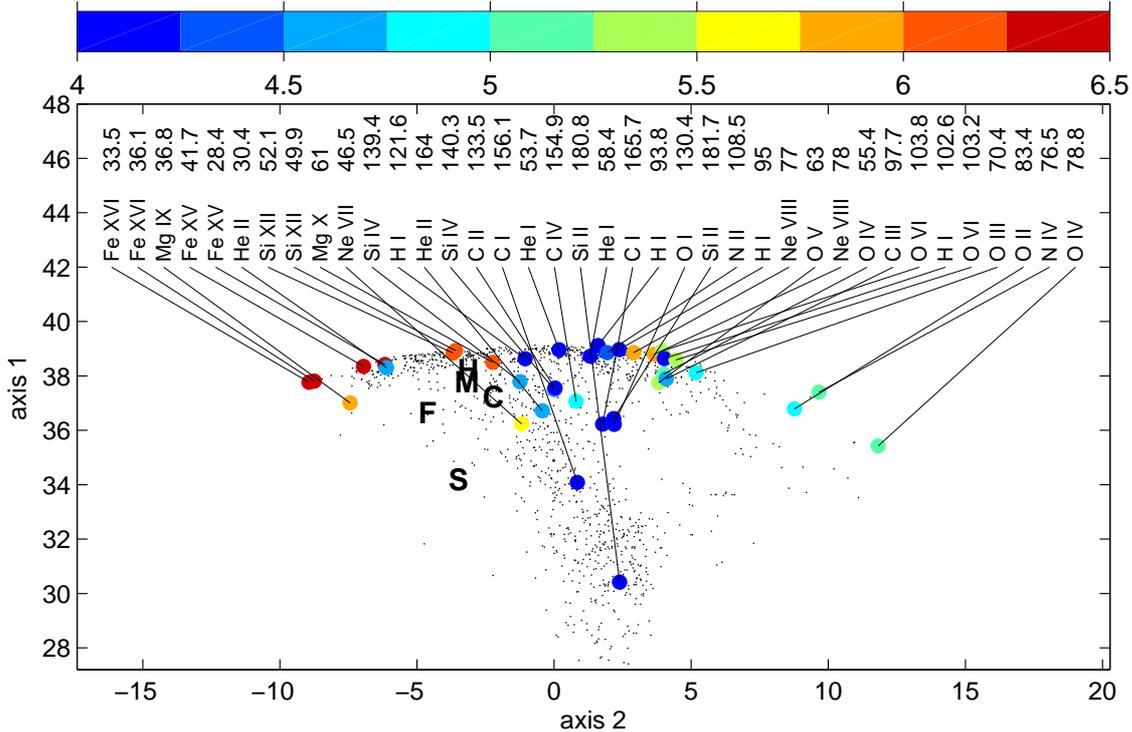}
\caption{Same connectivity map as in Fig.~\ref{fig_map_norm}, but with standardised data. Axes 1 is now the vertical one.}
\label{fig_map_stand}
\end{center}
\end{figure*}

The connectivity map we obtain with standardised data (Fig.~\ref{fig_map_stand}) differs quite substantially from the one obtained with normalised data. The curved boundary of the cluster of irradiances is a consequence of the standardisation, which distributes the points on a hypersphere in 42 dimensions, from which we see a two-dimensional projection. The spectral lines are now packed more tightly together, except for some cold and intense chromospheric lines that appear near the bottom. We believe this effect to be instrumental. The lowermost lines are indeed also strong ones that are more susceptible to degradation. Figure~\ref{fig_map_stand} must therefore be interpreted with care.

Notice that the difference between coronal and lower altitude lines is now much less evident than in Fig.~\ref{fig_map_norm}, even though a temperature ordering is still apparent with hot lines on the left, some transition region lines on the right and chromospheric lines in between. Unfortunately, none of the indices is located in the middle of any of these three clusters. Surprisingly, the three indices that were supposed to describe low altitude regions are now found to be much more closely associated with coronal lines. This had already been noticed before \citep{donnelly86,viereck01}. The Mg II index, for example, reproduces the He II line at 30.4 nm more effectively than the f10.7 index. All indices show about equal poor performance but the f10.7 index and the sunspot number are the least appropriate.

The representation we use in Figs.~\ref{fig_map_norm} and \ref{fig_map_stand} has the advantage of showing in a single glance what the relative correspondence between spectral lines and indices is. This representation confirms older results but also reveals convincingly why and by how much some indices fail to describe the EUV irradiance. Notice in particular that none of the indices is located right within a cluster of spectral lines. We conclude that none of them can fully reproduce the irradiance either with a linear relationship or with a direct proportionality. The key result is that EUV indices such as Mg II, Ca K and He I are appropriate for describing photospheric and chromospheric lines as far as their level of variability is concerned, but that their relative variations are better for reproducing coronal lines when it comes to short-term variations. This result is an incentive for decomposing EUV indices into different temporal scales before trying to reconstruct the EUV irradiance from them. 

Another interesting property of Figs.~\ref{fig_map_norm} and \ref{fig_map_stand} is worth mentioning: the connectivity map preserves distances. Any linear combination of two quantities will therefore be approximately located on a straight line that connects the two. Figure~\ref{fig_map_norm}, for example, shows that several spectral lines can be reconstructed from a linear combination of the f10.7 and Mg II (or Ca K) indices. Some could also be reconstructible using either He I and f10.7, or He I and Mg II (or Ca K). The Mg II and Ca K pair, however, or any combination containing the sunspot number, would be worthless.

Ideally, the indices should be located in different parts of the cluster of spectral lines (for both normalisations), in order for all spectral lines to be reconstructed effectively. This is in contradiction with the distributions we observe, in which the indices are either grouped together (with standardised data) or tend to be aligned (with normalised data). Figures~\ref{fig_map_norm} and \ref{fig_map_stand} thus vividly illustrate the EUV reconstruction problem by showing that no significant improvement can be expected from a linear combination of indices. As long as we do not find indices that cover (even partly) the different parts of the clusters of spectral lines, any irradiance reconstruction procedure that is based only on indices is deemed to miss significant features of the EUV variability.


\section{Conclusion}

The first objective of this study was to find a simple way of determining how well EUV spectral lines can be reconstructed from various EUV proxies.  We introduced a novel representation that expresses the measure of relatedness between any pair of spectral lines or proxies; 38 EUV lines between 26 and 194 nm and five EUV/UV indices (the sunspot number, the f10.7, Ca K, and Mg II indices, and the He I equivalent width) were compared. This representation readily shows why some indices are more adequate for reconstructing spectral lines that originate from one given solar altitude.

The second objective was to determine which indices would be needed for empirical modelling of spectral lines. It is difficult to give clear preference to one particular index (excluding instrumental constraints), apart from the sunspot number, which is the least appropriate by our standards. No single index can successfully describe both the level of variability on different time scales. The Mg II and the Ca K indices are appropriate for describing the long-term ($\gg 27$ days) evolution of the least-energetic part of the EUV spectrum but less so for modelling the short-term evolution. Conversely, the Mg II, Ca K and He I indices are found to be rather good proxies of the short-term evolution of coronal lines. No combination of indices was found to be appropriate for reconstructing different parts of the EUV spectrum. 

The next obvious step consists in doing a multiscale analysis and repeat this procedure after decomposing the data beforehand into different time scales. Another improvement, which is in progress, consists in using a more general measure of correlation that also gauges nonlinear dependencies.


\section*{Acknowledgements}

We gratefully acknowledge the TIMED/SEE team for providing the EUV irradiance data, the Royal Observatory of Belgium for the sunspot data, the Space Environment Center (NOAA, Boulder) for the f10.7 and Mg II indices, and the National Solar Observatory at Kitt Peak for the He I and Ca K data. This work was supported by COST action 724 and by the french solar-terrestrial physics programme.

\end{document}